\def\papertitle{Audio-visual talker localization in video for spatial sound reproduction}
\def\paperauthorA{Davide Berghi}
\def\paperauthorB{Philip J. B. Jackson}
\documentclass[twoside,a4paper]{article}
\usepackage{etoolbox}
\usepackage{dafx_24}
\usepackage{amsmath,amssymb,amsfonts,amsthm}
\usepackage{euscript}
\usepackage[T1]{fontenc}
\usepackage[utf8]{inputenc}
\usepackage{ifpdf}
\usepackage[english]{babel}
\usepackage{caption}
\usepackage{subfig} % or can use subcaption package
\usepackage{soul,color}
\usepackage{bbm}
\usepackage{siunitx}

\input glyphtounicode
\ninept

\newcounter{numauth}
\newcounter{listcnt}
\newcommand\authcnt[1]{\ifdefined#1 \stepcounter{numauth} \fi}

\newcommand\addauth[1]{
\ifdefined#1 
\stepcounter{listcnt}
\ifnum \value{listcnt}<\value{numauth}
\appto\authorslist{, #1}
\else
\appto\authorslist{~and~#1}
\fi
\fi}
\authcnt{\paperauthorB}
\authcnt{\paperauthorC}
\authcnt{\paperauthorD}
\authcnt{\paperauthorE}
\authcnt{\paperauthorF}
\authcnt{\paperauthorG}
\authcnt{\paperauthorH}
\authcnt{\paperauthorI}
\authcnt{\paperauthorJ}
\def\authorslist{\paperauthorA}
\addauth{\paperauthorB}
\addauth{\paperauthorC}
\addauth{\paperauthorD}
\addauth{\paperauthorE}
\addauth{\paperauthorF}
\addauth{\paperauthorG}
\addauth{\paperauthorH}
\addauth{\paperauthorI}
\addauth{\paperauthorJ}
\usepackage{times}
\newif\ifpdf
\ifx\pdfoutput\relax
\else
   \ifcase\pdfoutput
      \pdffalse
   \else
      \pdftrue
   \fi
\fi

\ifpdf % compiling with pdflatex
  \usepackage[pdftex,
    pdftitle={\papertitle},
    pdfauthor={\authorslist},
    pdfsubject={Proceedings of the 27th International Conference on Digital Audio Effects (DAFx24)},
    colorlinks=false, % links are activated as color boxes instead of color text
    bookmarksnumbered, % use section numbers with bookmarks
    pdfstartview=XYZ % start with zoom=100% instead of full screen; especially useful if working with a big screen :-)
  ]{hyperref}
  \usepackage[pdftex]{graphicx}
\else % compiling with latex
  \usepackage[dvips]{epsfig,graphicx}
  \usepackage[dvips,
    pdftitle={\papertitle},
    pdfauthor={\authorslist},
    pdfsubject={Proceedings of the 27th International Conference on Digital Audio Effects (DAFx24)},
    colorlinks=false, % no color links
    bookmarksnumbered, % use section numbers with bookmarks
    pdfstartview=XYZ % start with zoom=100% instead of full screen
  ]{hyperref}
\fi
\usepackage[hypcap=true]{caption}
\def\etal{\emph{et al}.\ }
\title{\papertitle}

\affiliation{
\paperauthorA\ and \paperauthorB \,\thanks{\vspace{-3mm}}}
{\href{https://dafx24.surrey.ac.uk}{Centre for Vision, Speech and Signal Processing} \\ University of Surrey\\ Guildford, UK\\
{\tt \href{mailto:dafx24@surrey.ac.uk}{\{davide.berghi, p.jackson\}@surrey.ac.uk}}
}

\begin{document}
% more pdf-tex settings:
\ifpdf % used graphic file format for pdflatex
  \DeclareGraphicsExtensions{.png,.jpg,.pdf}
\else  % used graphic file format for latex
  \DeclareGraphicsExtensions{.eps}
\fi

%\makeatletter
%\pdfbookmark[0]{\@pdftitle}{title}
%\makeatother

\maketitle

\begin{abstract}
Object-based audio production requires the positional metadata to be defined for each point-source object, including the key elements in the foreground of the sound scene.
In many media production use cases, both cameras and microphones are employed to make recordings, and the human voice is often a key element. 
In this research, we detect and locate the active speaker in the video, facilitating the automatic extraction of the positional metadata of the talker relative to the camera’s reference frame. 
With the integration of the visual modality, this study expands upon our previous investigation focused solely on audio-based active speaker detection and localization.
%The proposed method leverage the spatial sound-field recorded by a microphone array alongside the visual display captured by camera. 
Our experiments compare conventional audio-visual approaches for active speaker detection that leverage monaural audio, our previous audio-only method that leverages multichannel recordings from a microphone array, and a novel audio-visual approach integrating vision and multichannel audio.
We found the role of the two modalities to complement each other. Multichannel audio, overcoming the problem of visual occlusions, provides a double-digit reduction in detection error compared to audio-visual methods with single-channel audio. The combination of multichannel audio and vision further enhances spatial accuracy, leading to a four-percentage point increase in F1 score on the Tragic Talkers dataset.
Future investigations will assess the robustness of the model in noisy and highly reverberant environments, as well as tackle the problem of off-screen speakers. 

\end{abstract}

\section{Introduction}
\label{sec:intro}
%Immersive audio-visual production seeks to deliver

In 3D audio-visual production, meticulous attention is required to accurately align sound sources with the visual events they accompany. 
To produce and author an immersive experience, audio sources are generally treated as objects and manually placed in the virtual space by the producer.
This approach to production is often referred to as object-based media (OBM) production \cite{Coleman:2018:objectBased,pike2016object-based}.
In OBM, each object, spanning audio, video, graphics, text, or other forms of media, is accompanied by its metadata. The metadata associated with an individual object describes specific attributes or desired behaviors of the object, such as its content or position in space over time. 
OBM is valued for its adaptability and interactivity, allowing for tailored experiences based on user preferences or device configuration. For example, in the context of producing immersive spatial audio, object-based audio can theoretically be rendered on any loudspeaker configuration, unlike traditional channel-based mixes, which lack such flexibility.

\begin{figure}[tb]
\centerline{\includegraphics[width=\columnwidth]{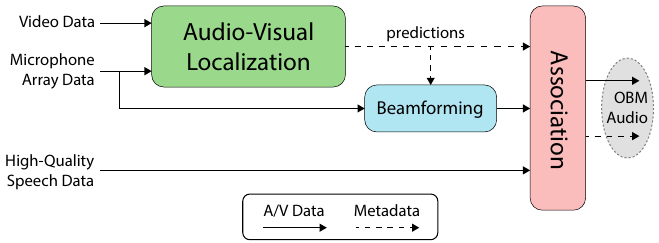}}
\caption{Pipeline for speech signals objectification proposed by Mohd Izhar \etal \cite{izhar:2020:AVtracker} and Schweiger \etal \cite{Schweiger:2022:tool6dof}. Positional metadata are automatically predicted by leveraging video and microphone array data. These predictions are not only used as final positional information for the spatialization of the objects but also to drive a spatial beamformer. Filtered signals extracted with the beamformer are associated with, and replaced by, the high-quality speech data recorded with Lavalier microphones. This paper focuses on the audio-visual prediction of the speaker's positional metadata.}
\label{fig:OBMpipe}
\end{figure}

However, while it is relatively trivial to manually spatialize synthetic audio-visual assets, this is not true for real scenes. Ideally, the ultimate spatialization should truthfully reflect the positioning and movements of the events. An example of a common audio-visual event is a moving talker. 
To tackle this problem, Mohd Izhar \etal \cite{izhar:2020:AVtracker} and Schweiger \etal \cite{Schweiger:2022:tool6dof} proposed the speaker objectification pipeline depicted in Figure \ref{fig:OBMpipe}.
Automated predictions regarding the locations of active speakers are generated using an audio-visual tracker, which utilizes both video data and audio captured through a microphone array. These predictions drive the spatial listening directivity of a beamformer employed to filter the signals from the microphone array. The resulting filtered speech signals are then associated with, and substituted by, the high-quality speech recordings obtained through Lavalier microphones.

This paper addresses the automatic extraction of the speaker positional metadata, corresponding to the green block of the pipeline presented in Figure \ref{fig:OBMpipe}. Specifically,
we focus on video-based horizontal active speaker detection and localization (ASDL) as videos are a widely consumed form of media, and subjective studies suggest that humans tend to be more spatially sensitive across the azimuth direction than elevation \cite{Strybel:2000:MinimumAA}.
The audio-visual localization is performed on the Tragic Talkers dataset \cite{Berghi:2022:TragicTalkers} and it expands upon our previous audio-only studies \cite{Berghi:2024:TASLP,Berghi:2021:mmsp}. Our previous works highlighted the benefits of employing multichannel audio captured with a microphone array, outperforming traditional video-based audio-visual approaches for active speaker detection that employ monaural audio and rely on visual face detection. 
Multichannel audio provides a higher detection accuracy as it does not suffer from visual occlusions. This paper demonstrates that partnering multichannel audio with the visual modality will improve spatial accuracy while preserving the high detection rate enabled by multichannel audio.

The remainder of this document provides an overview of the related work on video-based active speaker detection and localization, describes the proposed method and the audio-visual network architecture, presents and discusses our experimental results, concludes the paper and suggests future directions for the field.

\section{Related Work}

ASDL can be addressed through two distinct phases. Initially, the localization subtask is undertaken, wherein a visual face detector is utilized to pre-select a set of candidate speakers. Subsequently, the detected faces undergo classification into active or inactive. In computer vision, this second classification process is usually referred to as Active Speaker Detection (ASD) \cite{Roth:2020:AVA,zhang2021unicon,tao2021TalkNet,Alcazar_2020_CVPR,Chakravarty2016ActiveSD}.
Researchers usually partner the video stream with the respective (mono) audio signal. 
Pioneering the work on video-based active speaker detection, Cutler \etal \cite{Cutler:2000:lookWho} proposed to observe the correlation between mouth motion and audio data. %However, their method is based on the assumption that a single speaker is active at a time and minimum background noise is present. Furthermore, the head of the speaker is assumed not to move considerably during talking. 
%Nevertheless, this work inspired successive research in the field of voice activity detection \cite{Sodoyer:2006:AVVAD,takeuchi09_avsp}. 
%Sodoyer \etal \cite{Sodoyer:2006:AVVAD} leveraged visual speech information, namely lip movements, to improve the \ac{vad} performance in noisy environments. %Similarly, Takeuchi \etal \cite{takeuchi09_avsp} performed \ac{avvad} leveraging optical flow features extracted from lip images. Audio and visual features are then fused in a multi-stream \ac{hmm} for the final prediction.
Haider \etal \cite{Haider:2012:towardsspeaker} combined lip tracking and voice activity detection (VAD) to predict who is speaking in multiparty dialogue videos.
Chakravarty \etal \cite{Chakravarty2015WhosSA} proposed to also include head and upper-body motion as additional visual cues to detect the active speaker. They adopted a self-supervised solution to perform ASD by training a visual network under the supervision of its audio counterpart. 
Subsequent works \cite{Chakravarty2016ActiveSD,Hoover2019ICASSP}, adopted a two-step audio-visual co-training for speaker detection and identification. %A \ac{vad} is initially employed to automatically supervise the training of a personalised video-based active speaker classifier. Secondly, the video classifier is used to create a voice model for each person.
%A similar approach was adopted by Hoover \etal \cite{Hoover2019ICASSP}. 
They exploited the co-occurrence of speech and faces in videos to associate clusters generated from speech features with clusters generated from facial features.
%A facial identification system is adopted to extract the faces from the video frames before being grouped into clusters. A diarization system is then employed to divide the audio input into segments and a \ac{vlad} encoder is used to extract features from the speech segments which are then clustered. 
%Finally, leveraging statistics of speech and facial clusters across a video, each face is associated with the respective voice.
With the advance of deep learning techniques and the availability of larger datasets \cite{nagrani:2020:voxceleb,Chung16lipreading}, different yet related audio-visual tasks have been introduced, such as audio-visual lip reading \cite{Chung16lipreading,afouras:2020:ASR}, lip-voice synchronization \cite{Chung2018PerfectMI}, and audio-visual speaker separation \cite{Afouras18:theconv}.
What is probably the first, large, annotated dataset for ASD was released for the ActivityNet Challenge (Task B) at CVPR 2019: the AVA-ActiveSpeaker dataset \cite{Roth:2020:AVA}. 
It provides 38.5 hours of audio-visual face tracks (sequences of consecutive face crops) labeled for speech activity.
Since the face tracks are provided, the challenge task consists of classifying each face as active or inactive, leveraging audio and video signals. 
%Chung \textit{et al.} \cite{Chung2019NaverAA} and Zhang \etal \cite{Zhang2019MultiTaskLF} achieved respectively the first and the second positions at the 2019 challenge. Both their approaches applied 3D convolutions on the audio and visual signals 
At the 2019 challenge, the first \cite{Chung2019NaverAA} and second \cite{Zhang2019MultiTaskLF} positions were achieved by leveraging 3D convolutional neural networks (CNNs).
%, employing models such as the 3D-ResNet18 \cite{Stafylakis:2017:CombiningRN} and the VGG-M network \cite{chatfield2014returnDevil} pre-trained on audio-to-video synchronisation \cite{Chung16:outof,Chung2018PerfectMI}.
After that, Alcázar \textit{et al.} proposed a model called Active Speaker in Context (ASC) \cite{Alcazar_2020_CVPR}. Instead of compute-intensive 3D convolutions or large-scale audio-visual pre-training, ASC leverages context: in assessing the speech activity of a candidate speaker, it looks at any other available faces. %The model is composed of three blocks: a short-term encoder to extract audio-visual features from the frames, a pairwise attention refinement to better model multi-speaker scenarios, and a temporal refinement to observe long-time horizons.
Zhang \textit{et al.} \cite{zhang2021unicon} also tackled the ASD task by leveraging contextual information and proposed the UniCon network. 
Tao \textit{et al.} introduced TalkNet \cite{tao2021TalkNet}, an ASD model that leverages short- and long-term features. %Inter-modal temporal encoders extract feature embeddings from the respective modalities, then, audio-visual cross-attention is employed to fuse the two embeddings. Finally, a self-attention block is used to capture long-term speaking evidence.
Additionally, motivated by the call for an ASD system that works properly outside the AVA-ActiveSpeaker dataset domain, they formed a second ASD dataset based on LRS3 \cite{Afouras:2018:LRS3} and VoxCeleb2 \cite{Chung:2018:voxceleb2} called TalkSet \cite{tao2021TalkNet}. 
Recently, Alcázar \textit{et al.} \cite{Alcazar2022EndtoEndAS} proposed an end-to-end ASD that unifies audio-visual feature extraction and spatio-temporal context aggregation. %To reduce the computational cost that characterises most of the existing \ac{asd} models, Liao \etal \cite{Liao_2023_lightASD} presented a light-weight architecture with an audio-visual decoder based on \ac{gru}. 

However, these ASD solutions, focusing solely on the audio-visual classification of the provided face tracks, depend on visual face detection for the localization subtask.
In practice, the speaker can be occluded or facing away from the camera, leading to face detection failures and subsequent degradation of the overall system performance. When this happens, monaural audio is insufficient to compensate for the visual failure, as it lacks the necessary spatial cues required to locate audio sources. In other words, the active speaker is detected only when visible.
In previous studies \cite{Berghi:2021:mmsp,Berghi:2024:TASLP}, we overcame this problem by leveraging multichannel audio to simultaneously address the detection and localization aspects of ASDL. Our model was able to locate the active speaker in the video frames solely from audio inputs, achieving better detection and recall rates.

Some other recent studies partnered multichannel audio with vision for audio-visual ASDL. For example, Qian \textit{et al.} in \cite{Qian:2021:AVFusion} and \cite{Xinyuan:2023:AVcrossAtt} employed visual feature vectors encoding face bounding box coordinates to improve spatial accuracy. The detected face bounding boxes are represented as horizontal and vertical Gaussian-like vectors. %In this representation, the peaks of the Gaussians correspond to the central coordinates of the bounding box, as depicted in Figure \ref{fig:lit:Gauss} for the single speaker scenario. This representation can be extended to multiple speakers by incorporating the sum of multiple Gaussians in the vector representation.
%A \ac{cmaf} mechanism \cite{Xinyuan:2023:AVcrossAtt} which includes cross-attention and self-attention modules is proposed to integrate the features extracted from the audio and the visual modalities. 
Similarly, Wu \etal \cite{Wu:2023:multiSpeakDOA} adopted Gaussian-like vectors as visual input features and a transformer-based network \cite{Vaswani:2017:AttAllUNeed} for their predictions. %As audio input features, the researchers employed complex multichannel spectrograms.
Zhao \etal \cite{Zhao:2022:visAss} proposed a self-supervised student-teacher knowledge distillation approach to train a multichannel audio network from visual supervision.
%Li \etal \cite{Li:2022:multiModPerc} leveraged the camera model to align \ac{gcf} acoustic maps to the image frame localisation space. 
In a different study, they explored the audio-visual speaker localization from egocentric views \cite{zhao:2023:AVspeakEgo}. That is, the prediction is performed from the point of view of the device ``wearer'', who is free to move in space and the sensors are therefore not stationary.
Research groups from Meta Reality Labs released the EasyCom dataset \cite{donley2021easycom}. It consists of a set of videos captured from custom AR glasses with a camera and a microphone array integrated.
Jiang \textit{et al.} \cite{jiang2022egocentric} employed EasyCom to perform audio-visual speaker localization in an egocentric setting.
Concurrently, Gurvich \etal \cite{gurvich2023realtime} performed the real-time ASDL with a microphone array in 360° videos.
Similarly, in this paper, we integrate multichannel audio and visual data. However, instead of leveraging visual features such as Gaussian-like vectors, we focus on the integration of audio and visual embeddings extracted with pre-trained audio and visual encoders \cite{Berghi:2024:ICASSP24}.

\section{Method}

The proposed method extracts audio and visual feature embeddings with an audio and a visual encoder, respectively. The embeddings are then concatenated and fed to an attention-based unit. The output of the attention-based unit represents a joint latent audio-visual representation of the input signals, which is used to generate the final prediction through a feed-forward network. A schematic representation of the proposed model is presented in Figure\,\ref{fig:architecture}.
%By processing video and multichannel audio data, this model capitalizes on both modalities

\subsection{Dataset} 

\begin{figure}[tb]
\centerline{\includegraphics[width=\columnwidth]{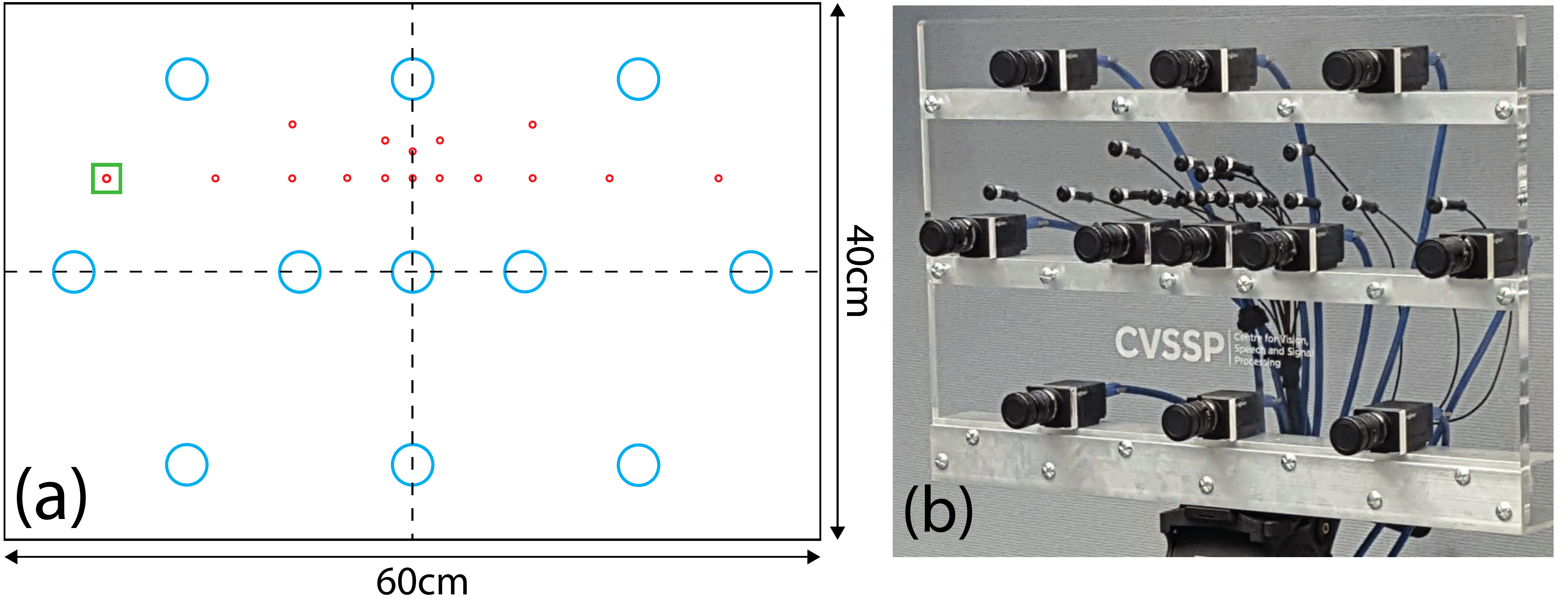}}
\caption{(a) Schematic of camera (blue circles) and microphone (red dots) positions on the AVA Rig. The green square highlights the reference microphone. (b) Photo of an AVA Rig.}
\label{fig:AVArig}
\end{figure}

To train and evaluate the proposed method, we employed the Tragic Talkers dataset \cite{Berghi:2022:TragicTalkers}. 
It offers sequences captured by two Audio-Visual Array (AVA) Rigs. Each AVA Rig is a light-field and sound-field sensing platform consisting of a 16-element microphone array and 11 cameras fixed on a flat perspex baffle as depicted in Figure\,\ref{fig:AVArig}. Therefore, each sensor is located in a fixed relative position and orientation with respect to the other sensors. 
The microphone array has a horizontal aperture of 450\,mm and a vertical aperture of only 40\,mm. This results in higher resolution when localizing audio sources along the azimuth direction than elevation, which is consistent with human perception \cite{Strybel:2000:MinimumAA}.
Horizontally, the microphones are log-spaced for broad frequency coverage from 500\,Hz to 8\,kHz, to better support the horizontal speech band resolution.
Tragic Talkers was captured in an acoustically treated studio with an average reverberation time of 0.3s in the mid 0.5-2\,kHz frequency range and minimal background noise floor (SNR\,$\geq$\,30\,dB).
The dataset does not contain sequences in which the speakers talk simultaneously, off-screen talkers, or external sources of sound other than speech, making it ideal for audio-visual speaker diarization applications too. 
 %This makes the task easier, yet provides the groundwork to build a deep understanding of the case study and evaluate the proposed learning method. 
%Studying at most one active talker allows rigorous assessment of the proposed learning method in a studio environment as a realistic media production setting.
Tragic Talkers was specifically designed for OBM production research. 
Its content allows the implementation of the speaker objectification pipeline presented in Figure \ref{fig:OBMpipe} \cite{izhar:2020:AVtracker,Schweiger:2022:tool6dof}. In fact, it provides high-quality speech signals recorded with Lavalier microphones. Additionally, the scenes are captured against a blue background that facilitates the actors' silhouette extraction, empowering the producer to extract and position the actors as desired while retaining the natural motion of the performance.

The method for ASDL proposed in this paper necessitates a single video feed with the microphone array. 
Leveraging the multiple views available enables us to extend the network training by choosing the relevant camera perspective. 
Consequently, a one-hot vector denoting the selected view is appended to the input data, augmenting the dataset with diverse camera perspectives and enabling the network to learn the correct mapping to the desired viewpoint.
The sequences include one or two actors positioned at a distance of about 3--4\,m, engaging in monologues, conversations, and interactive scenes involving movement and occlusion.
Tragic Talkers comprises 30 scenes captured with two AVA Rigs. 
Each rig's audio-visual stream is employed independently, i.e., 16-channel audio is used to predict the speaker's position within any of the 11 camera perspectives of the rig. 
So the dataset's 30 scenes provide 60 rig sequences, each offering 11 viewpoints. 
The dataset is partitioned into a 50-sequence development set and a 10-sequence test set.
TragicTalkers provides ground truth (GT) labels for voice activity and 2D face bounding box. The test set used for evaluation also includes 3D mouth coordinates.

\subsection{Network Architecture}

\begin{figure}[tb]
\centerline{\includegraphics[width=0.7\columnwidth]{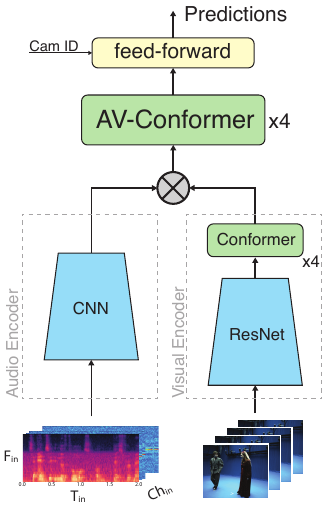}}
\caption{Proposed network architecture for audio-visual ASDL. An audio encoder based on a CNN extracts an audio embedding from the audio input features. Similarly, an encoder consisting of ResNet50 \cite{He:2016:resnet} followed by a Conformer unit \cite{Gulati2020ConformerCT} extracts a visual embedding from the video frames. $\otimes$ denotes the concatenation operation. After concatenation, the audio-visual features and processed by a second Conformer unit. A feed-forward network generates the final prediction. A camera ID one-hot vector is used to regress the speaker's position to the desired camera view.}
\label{fig:architecture}
\end{figure}

We tested the proposed architecture in a recent study where we tackled the sound event localization and detection (SELD) task \cite{Berghi:2024:ICASSP24}. Here, we aim to explore whether a similar network can be extended to the more specific ASDL tasks too.  
As depicted in Figure\,\ref{fig:architecture}, the network presents an audio and a visual encoder to extract audio and visual feature embeddings. The embeddings are then concatenated and fed to an audio-visual Conformer \cite{Gulati2020ConformerCT}. Conformer models were originally proposed for speech recognition but lately, they have achieved state-of-the-art performance in tasks such as SELD too \cite{Wang:2023:ACS}. They present an architecture similar to the Transformer proposed by Vaswani \etal \cite{Vaswani:2017:AttAllUNeed}, however, they integrate a convolution module that performs pointwise and depthwise convolutions.
Inspired by Wang \etal \cite{Wang:2023:ACS}, 4 layers are employed with 8 heads each. The size of the kernel for the depthwise convolutions is set to 51, and the dimension of the hidden layer in the feed-forward networks of each Conformer layer is 1024 \cite{Wang:2023:ACS}. 
The output of the Conformer, i.e., the latent audio-visual representation, is then fed to a feed-forward unit consisting of three fully-connected layers to make the final predictions. A one-hot vector encoding the desired camera view of the AVA Rig is concatenated to the latent audio-visual vector after the second fully-connected layer. It allows mapping the final prediction to the desired camera of the rig.
In output, the network predicts the speaker's horizontal location and voice activity confidence at a temporal resolution that matches the label and video frame rate (i.e., a position-confidence pair is generated for each video frame).

\subsubsection{Audio Encoder}

The audio encoder takes as input spatial features extracted from the multichannel audio signals (a detailed description of the audio input features will be provided in \ref{sec:inputs}) with shape $Ch_{in}\times T_{in}\times F_{in}$, where $Ch_{in}$ corresponds to the number of channels of the input spatial features, $T_{in}$ the number of temporal bins, and $F_{in}$ the frequency bins.
The audio encoder presents a CNN architecture. The CNN architecture consists of four convolutional blocks, each consisting of two 3$\times$3 convolutional layers followed by average pooling, batch normalization \cite{ioffe:2015:BN}, and ReLU activation. The average pooling layer is applied with a stride of 2, halving the temporal and frequency dimension at each block. The resulting tensor of shape $512\times T_{in}/16\times F_{in}/16$ is then reshaped and frequency average pooling is applied to achieve a $T_{in}/16\times 512$ dimensional feature embedding. $T_{in}$ is chosen so that $T_{in}/16$ matches the label frame rate of the Tragic Talkers dataset (30 labels per second).
%The only improvement with the CNN backbone used in the previous chapters consists in the introduction of residual connections \cite{He:2016:resnet} at each convolutional block as they provided a marginal increment in performance during preliminary tests.

\subsubsection{Visual Encoder}

As a visual encoder, the ResNet50 model \cite{He:2016:resnet} followed by a Conformer \cite{Gulati2020ConformerCT} is tested.
Each video frame is fed to ResNet50. Since the video streams of Tragic Talkers are captured at 30\,fps, ResNet50 extracts a number of frame embeddings that match the label frame rate as well as the audio embedding temporal resolution. %The ResNet50 model of the torchvision library was employed, which is pre-trained on ImageNet \cite{ImageNet:2009:dataset}. 
The frame embeddings extracted by ResNet50 are further processed by the Conformer module, which presents the same hyper-parameters utilized for the one employed for the audio-visual fusion.
%Additionally, to take full advantage of the video temporal dimension, a second visual encoder based on 3D CNN was tested. The I3D model by Carreira \etal \cite{Carreira2017QuoVA} was adopted as it is a widely-employed yet effective model for action recognition and 3D feature extraction. %I3D is pre-trained on Kinetics \cite{Kay2017kinetics}. The original model comprises two branches: one for the RGB video and one for the optical flow frames. For simplicity, only the RGB branch is employed in this study. 
%For computational reasons, the input frame rate adopted with the ResNet-Conformer (10fps) encoder was kept with I3D. 3D CNNs have the effect of downsampling the temporal dimension. Therefore, the output of I3D is then temporally interpolated to reconstruct the label frame rate. 
The video frames used as inputs to the visual encoders are resized to 224x224p. We employed the ResNet50 model available with the torchvision library\footnote{\href{https://pytorch.org/vision/stable/index.html}{https://pytorch.org/vision/stable/index.html}}, which is pre-trained on image classification on the ImageNet dataset \cite{ImageNet:2009:dataset}.
Before being fed to the Conformer, the frame embeddings are resized from the original 2048-dimensional vectors generated from ResNet50 to 512 dimensions employing a fully-connected layer to match the size of the audio embeddings.

The remainder of this paper will often refer to the visual encoder as ResNet-Conformer and to the attention-based audio-visual fusion as AV-Conformer.

\subsection{Audio Input Feature} \label{sec:inputs}

This work adopts log-mel spectrograms concatenated with generalized cross-correlation with phase transform (GCC-PHAT) features in log-mel space \cite{Knapp:gccphat:1976,Cao:2019:polyphonic} extracted from the microphone array signals. These audio features were chosen because they achieved good performance and robustness on the large microphone array of the Tragic Talker dataset \cite{Berghi:2023:WASPAA}.

The GCC-PHAT is employed to estimate the time difference of arrival (TDOA) of a sound source at two microphones \cite{Knapp:gccphat:1976}. The idea is to find the lag time that maximizes the cross-correlation function between the signals sensed by the two microphones. The generalized cross-correlation (GCC) is computed through the inverse Fast-Fourier Transform (inverse-FFT) of their cross-power spectrum. Phase-transformed GCC, namely the GCC-PHAT, eliminates the influence of the amplitude by leaving only the phase \cite{Cao:2019:polyphonic}. 
The GCC-PHAT between the \textit{i}-th and the \textit{j}-th microphone is defined at each audio frame \textit{t} as:
\begin{equation}
    \label{eq:gcc_phat}
    GCC_{ij}(t,\tau)=\mathcal{F}^{-1}_{f\rightarrow\tau} \frac{\mathbf{X}_i(t,f)\mathbf{X}^*_j(t,f)}{|\mathbf{X}_i(t,f)\mathbf{X}^*_j(t,f)|},
\end{equation}
where $\mathbf{X}_i(t,f)$ is the Short-Time Fourier Transform (STFT) of the \textit{i}-th channel, $\mathcal{F}^{-1}_{f\rightarrow\tau}$ the inverse-FFT from the frequency domain $f$ to the lag-time domain $\tau$, and $(.)^*$ denotes the complex conjugate. The TDOA can be estimated as the lag-time $\Delta\tau$ that maximizes $GCC_{ij}(t,\tau)$. 

With time bins ($t$) on the $x$-axis and time-lags ($\tau$) on the $y$-axis, the GCC-PHAT can be concatenated with the log-mel spectrograms extracted from the channels of the microphone array, as indicated by Cao \etal \cite{Cao:2019:polyphonic}. The concatenation of GCC-PHAT features and log-mel spectrograms provides the network with a unified representation that enables both detection and localization subtasks. We compute the GCC-PHAT between a reference microphone and the other microphones of the array. As the reference microphone, we select the first channel of the lower sub-array, as highlighted in Figure\,\ref{fig:AVArig} (a). From the same channel, we also extract a single log-mel spectrogram for the concatenation.

\subsection{Loss Function}\label{ASDL:met:loss}

For each input segment, the loss function $\mathcal{L}$ is determined as the sum of the individual frame losses. A sum-squared error loss \cite{Redmon:2016:YOLO} is computed at each output frame and is comprised of a regression and a voice activity confidence loss:
\begin{equation}
\mathcal{L}=\sum_{i=1}^{T_{in}/16}\mathbbm{1}_i(x_i-\hat{x}_i)^2+(C_i-\hat{C}_i)^2
\label{loss_function}
\end{equation}
where $x_i$ and $\hat{x}_i$ are respectively the predicted and target positions of the speaker along the horizontal axis of the $i$-th video frame, while $C_i$ and $\hat{C}_i$ are the predicted and target confidences. The voice activity confidence loss is trivially achieved using the voice activity annotations: $\hat{C}_i$ is set to 1 when the frame is active and 0 when silent.
The masking term $\mathbbm{1}_i$ is 1 only when voice activity GT is true. It is set to 0 otherwise. So, when the frame is silent, the network is only penalized by the voice activity confidence loss and not by the regression loss. The target position of the speaker, $\hat{x}_i$, corresponds to the horizontal position of the center of the face bounding box of the active speaker, normalized by the size of the video frame to be in the range [0, 1].

\section{Experiments}

\subsection{Implementation Details Evaluation Metrics}

The network is trained with a 5-fold cross-validation approach: each validation fold sets aside 10 unseen sequences from the 50 sequences of the development set. This cross-validation approach is used to find suitable hyper-parameters for the network. Once found, the model is retrained using the entire 50-sequence training set with these values.
The network is trained for 50 epochs using batches of 32 audio feature inputs and Adam optimizer. The learning rate is fixed for the first 30 epochs, then reduced by 10\% each epoch, as in \cite{Cao:2019:polyphonic}. 
The initial learning rate determined in the cross-validation is \num{e-4}.

The audio stream of the Tragic Talkers dataset is sampled at 48 kHz. The dataset is discretized into audio-visual segments of 2 seconds. The label frame rate is consistent with the video frame rate (30 fps).
To align the output temporal resolution ($T_{out}=T_{in}/16$) with the labels frame rate, i.e., generating 60 activity-regression pair predictions for the 2-second input, an STFT with Hann window is applied at hop steps of 100 samples. Thus, the 2-second (96k-sample) audio chunk is discretized into 960 temporal bins (96k$/$100), which correspond to $T_{in}$. The Hann window presents a size of 512 samples, as in \cite{Nguyen:2021:SALSA}.
To compute the log-mel spectrogram used in the concatenation with the GCC-PHAT features, the frequency resolution of the spectrogram is down-sampled over 64 mel-frequency bins and the logarithm operation is applied. 
%Also, the distance between the two furthest microphones in the array, $d_{max}$, is 450 mm and the horizontal \ac{fov} of the camera, $\xi$, is roughly 55°. 
The number of time-lags for the GCC-PHAT is also set to 64 to enable the concatenation.
%For the log-linear spectrogram and the NIPD/IPD features used in SALSA-Lite and SALSA-IPD, due to the limited presence of speech information at higher frequencies and to avoid spatial aliasing, an upper cutoff frequency of 6 kHz is applied. 
%This extracts exactly the first 64 frequency bins and is therefore consistent with the input shape of GCC-PHAT. The logarithm operation is then applied.

The evaluation is performed on the TragicTalkers test set, labeled for 2D speaker mouth positions.
A frame prediction is considered positive, i.e. the network predicts the presence of speech, when the predicted voice activity confidence is above a threshold $Th$, and a positive detection is true when the localization error is within a predefined spatial tolerance $S$. 
The precision and recall rates are computed by varying the confidence threshold $Th$ from 0\% to 100\% sampling the thresholds from a Sigmoid-spaced distribution to provide more data points for high and low confidence values. 
The average precision (AP) was computed as the numerical integration of the precision-recall curve, as indicated in \cite{Everingham:2015:pacalVOC}. 
We set a spatial tolerance $S$ of $\pm$2° along the azimuth according to human auditory perception \cite{Strybel:2000:MinimumAA}, the minimum audible angle (MAA), corresponding to $\pm$89 pixels on the image plane. %\footnote{All conversions from pixels to degrees and vice-versa are performed with pre-computed camera-calibration data.}.  
From the precision and recall rates, the F1 score is computed too. %It represents a useful metric for comparing the different baselines employed, as it helps to find the optimal compromise between precision and recall rates and therefore provides an overall ASDL performance scale. 
To independently evaluate the localization and the speaker detection subtasks, we define the average distance (aD) and the detection error (Det Err \%) metrics. The former represents the average distance error between the active detections and the GT speaker locations. It is computed in pixels on the image frame and then converted to angle units leveraging the camera calibration data.
%The detection error is achieved by setting $Th=0.5$ to binarize as active or silent the voice activity confidence predictions. The percentage of incorrect predictions represents the detection error.
The detection error corresponds to the percentage of frames incorrectly classified as active or inactive when a threshold $Th=0.5$ is set.

%To run tests of statistical significance and standard errors for methods' comparison, the metrics were also separately computed for each test sequence in order to generate multiple data points.  
%Both aD and Det Err are first separately computed for each of the test sequences, then, the mean is taken. So, the results are reported with their respective standard error of the mean (Std Err).

\subsection{Methods and Baselines}

The proposed audio-visual method with multichannel audio is compared with two traditional audio-visual approaches for active speaker detection that employ single-channel audio and rely on face detection: Active Speaker in Context (ASC) \cite{Alcazar_2020_CVPR} and TalkNet \cite{tao2021TalkNet}. 
We also report the results achieved in our previous multichannel audio-only study \cite{Berghi:2024:TASLP}. The audio-only approach proposed in \cite{Berghi:2024:TASLP} leverages a convolutional recurrent neural network (CRNN) with an architecture similar to the audio encoder proposed in the present paper. However, instead of the Conformer unit, the CRNN presents two bidirectional gated recurrent units (biGRUs). 
Additionally, as a baseline system to highlight the advantages introduced by multichannel audio, we report the results achieved with a single-channel audio-only network (Mono). The input to the Mono baseline is a log mel spectrogram extracted from the central microphone of the array.

\subsection{Results}

\begin{table}[tb]
\caption{ASDL results on the test set of the Tragic Talkers dataset and modality potential. The results for the proposed approach are achieved with audio-visual inputs and multichannel audio (AV-M). The table includes the results achieved by a single-channel audio-only network (A-S), two audio-visual systems that employ single-channel audio (AV-S), and an audio-only method that leverages multichannel audio (A-M).
}
\begin{center}
%\footnotesize
\begin{tabular}{c|c|c|c|c|c}
\hline

\textbf{Method}&\textbf{Mod}&\textbf{DetErr}&\textbf{aD}&\textbf{AP}&\textbf{F1} \\ 
\hline
%\textsc{Asc} & AV-S & 43\% & 50p,\,1.1° & 59\% & 0.676 & 60\% & 0.679  %\\
%\textsc{Asc(s)} & AV-S & 24\% & \textbf{23p,\,0.52°} & 62\% & 0.704 & %63\% & 0.706  \\
Mono & A-S & \textbf{2.7\%} & 210p,\,4.7° & 10\% & 30.0 \\
ASC\,\cite{Alcazar_2020_CVPR} & AV-S & 43\% & 50p,\,1.1°& 59\% & 67.6 \\
TalkNet\cite{tao2021TalkNet} & AV-S & 14\% & 35p,\,0.79° & 82\% & 84.9 \\ 
%\hline
CRNN\,\cite{Berghi:2024:TASLP} & A-M & 3.2\% & 39p,\,0.88° & 87\% & 90.9 \\
\hline
\textbf{Proposed} & AV-M & \textbf{2.7\%} & \textbf{32p,\,0.72°} & \textbf{92\%} & \textbf{94.9} \\
\hline

\end{tabular}
\label{tab:results}
\end{center}
\end{table}

\begin{figure}[tb]
\centerline{\includegraphics[width=\columnwidth]{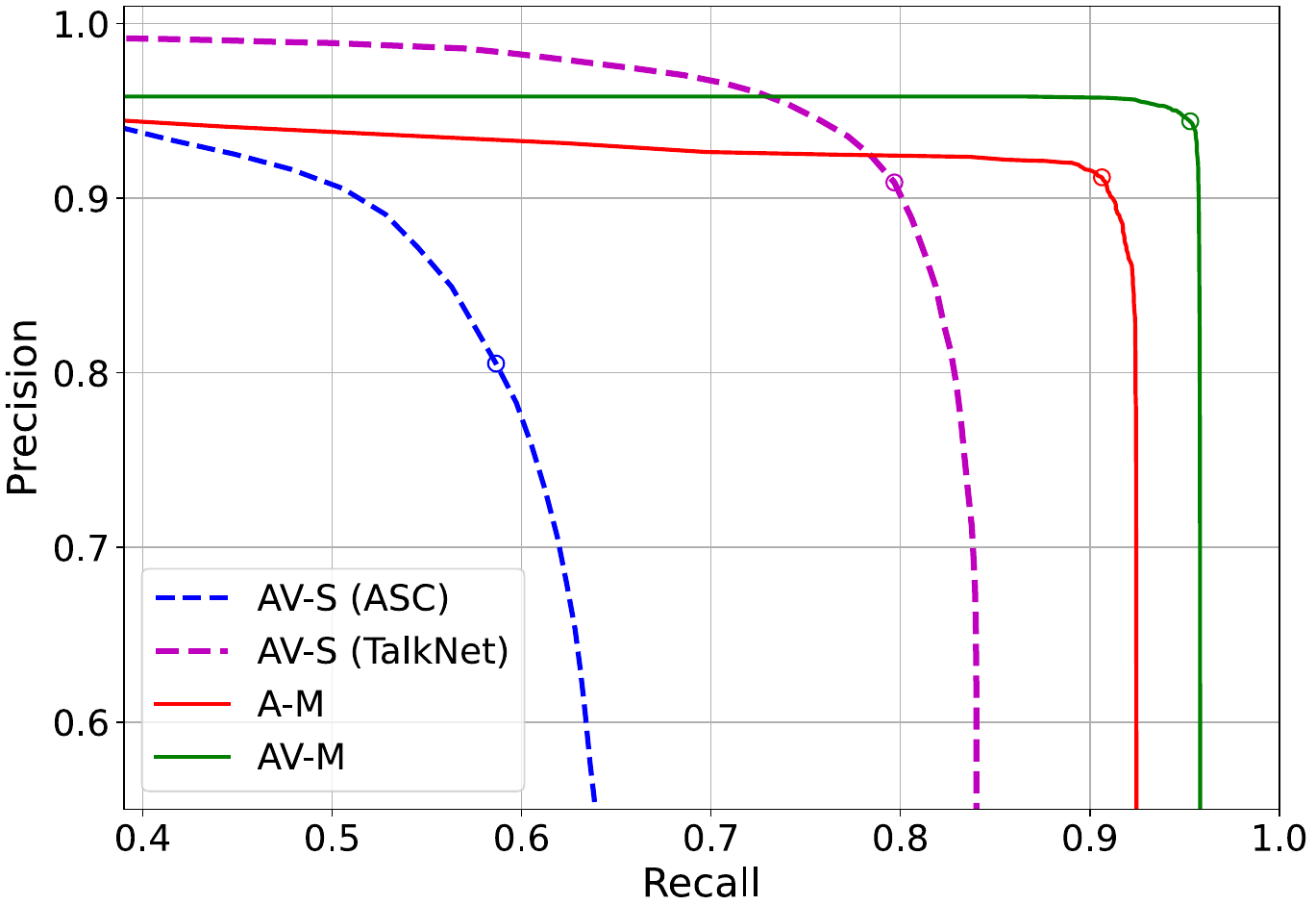}}
\caption{Comparison of precision versus recall curves for different ASDL methods. The plot includes audio-visual methods with single-channel audio (AV-S), i.e., ASC \cite{Alcazar_2020_CVPR} and TalkNet \cite{tao2021TalkNet}, the multichannel audio-only CRNN (M-S) \cite{Berghi:2024:TASLP}, and the proposed audio-visual approach with multichannel audio (AV-M). The combination of precision and recall rates that achieves the highest F1 score is marked on each curve.}
\label{fig:prec_rec}
\end{figure}

The experimental results are presented in Table \ref{tab:results}. 
The audio-visual multichannel (AV-M) method significantly improves the performance of the audio-only systems across nearly all metrics.

In the Mono baseline, the detection subtask is accomplished with a detection error of only 2.7\%. However, the position of the speaker is not accurately predicted due to the absence of spatial cues. To minimize the error, the model locates the speaker in the central area of the frame.

Conventional active speaker detectors, such as ASC \cite{Alcazar_2020_CVPR} and TalkNet \cite{tao2021TalkNet}, employ the audio modality only to classify the pre-extracted faces. The localisation subtask is performed by the visual face detector, yielding high spatial accuracy. However, since the average distance is achieved as the average of the predicted positions, the final average includes true and false positive predictions (false positive predictions happen when the detector classifies the silent actor as active).
%In other words, the aD values reported in Table \ref{tab:results} for ASC and TalkNet are influenced by the silent actor being misclassified as active. 
False positive detections are mainly present in the detections of ASC \cite{Alcazar_2020_CVPR}, penalizing its spatial accuracy as well as its detection.
In the audio-visual methods with single-channel audio, the horizontal coordinate of the center of the bounding box is used as prediction, while the ground truth used for evaluation refers to the actual mouth position of the speaker. Therefore, the aD achieved is slightly overestimated due to the offset between the two representations. For example, when the speaker is captured in profile and his/her mouth is closer to the edge of the bounding box. 
Additionally, the AV-S methods fail when the face of the active speaker is not detected by the face detector. This causes a higher detection error and, consequently, a lower recall rate, as shown in Figure\,\ref{fig:prec_rec}.
At its best precision-recall pair TalkNet presents a recall rate of 79.7\%, while ASC only of 58.5\%. As a consequence, their overall F1 scores are affected too. 
In contrast, the multichannel audio method achieves a lower detection error as speech activity can be sensed even when the speaker is visually occluded.
In fact, the double-digit detection errors of the AV-S methods are reduced to 3.2\% with the A-M approach. Figure\,\ref{fig:prec_rec} shows how the gap in recall rate generated by TalkNet is halved with the multichannel audio method (90.6\% recall rate).
This produces an AP and F1 score higher than the AV-S systems. %The F1 score achieved by GCC-PHAT (ref.) is significantly higher than the one achieved by TalkNet (p$=$0.02). The active frames that GCC-PHAT (ref.) and SALSA-Lite do not detect are mainly caused by wide predictions, where speech activity was correctly identified but the estimated locations were outside the tolerance. In fact, if the spatial threshold is set to the broader 5-degree tolerance,  the residual errors in terms of AP and F1 score for the GCC-PHAT (ref.) and SALSA-Lite methods are under 3\% and predominantly attributable to the detection errors.

When multichannel audio is partnered with vision, beneficial effects involve both detection and localization accuracy. 
The detection error decreases by 0.5 percentage points to 2.7\%, while the aD outperforms even the audio-visual TalkNet model. 
The F1 score is 4 percentage points higher than the A-M approach and 10 points greater than TalkNet.

The residual error in the F1 score for the proposed AV-M method is only 5\%. 
In the future, this error might be further narrowed by implementing visually guided predictions post-processing \cite{Wang:2023:dcase23}. For example, pose detection could rectify spatial predictions using the mouth key-point coordinates. This would further improve the localization accuracy and consequently increase the number of true positive detections that fall within the spatial tolerance.
Another aspect to consider is the consistency between training and testing labels. The GT labels employed to train the model correspond to center of the face bounding boxes, whereas the evaluation is based on GT mouth positions. This disparity introduces a subtle domain bias between the training and inference phases, potentially resulting in residual errors regardless of the quality of the model.

\section{Conclusions}

This paper proposes an audio-visual approach for active speaker detection and localization that leverages multichannel audio on the Tragic Talkers dataset. The approach extracts audio and visual embedding leveraging audio and visual encoders. Then, the embeddings are concatenated and processed by an AV-Conformer.
The proposed method outperforms conventional audio-visual approaches for active speaker detection that rely on visual face detection as well as our previous audio-only multichannel work. This highlights the importance of the input modalities.
Active speaker detection and localization can be employed for the automatic extraction of speaker positional metadata useful in immersive audio productions.

\section{Acknowledgments}
This research was funded by EPSRC-BBC Prosperity Partnership `{AI4ME}: Future personalised object-based media experiences delivered at scale anywhere' (EP/V038087/1). 
For the purpose of open access, the authors have applied a Creative Commons Attribution (CC BY) license to any Author Accepted Manuscript version arising. 
Data supporting this study are available from \url{https://cvssp.org/data/TragicTalkers}.

%\newpage
\nocite{*}
\bibliographystyle{IEEEbib}
\bibliography{DAFx24} % requires file DAFx24_tmpl.bib

\end{document}